\documentclass[aps,prl,10pt,showpacs,preprintnumbers,twocolumn,amsmath,amssymb,superscriptaddress,floatfix]{revtex4-1}
\usepackage[dvipdfmx]{graphicx}
\usepackage[dvipsnames]{xcolor}

\usepackage{bm}
\usepackage{textcomp}
\usepackage{units}
\usepackage{braket}

\usepackage[colorlinks,linkcolor=Blue,citecolor=Blue,urlcolor=Blue]{hyperref}
\newcommand{\figref}[2]{\ref{#1}\hyperref[#1]{#2}}

\makeatletter
\def\NAT@spacechar{}

\def\@hangfrom@section#1#2#3{\@hangfrom{#1#2}#3}
\def\@hangfroms@section#1#2{#1#2}

\newcommand{\secnumdot}{
  \setcounter{secnumdepth}{1}
  \renewcommand\@seccntformat[1]{\csname the##1\endcsname.\quad}
}

\newcommand{\beginmethods}{
  \par\bigskip
  \section*{Appendix}
  \setcounter{section}{0}
  \renewcommand{\thesection}{\arabic{section}}
  \secnumdot
}

\newcommand{\beginsuppinfo}{
  \par\bigskip
  \section*{Supplementary Information}
  \setcounter{section}{0}
  \setcounter{equation}{0}
  \setcounter{figure}{0}
  \setcounter{table}{0}
  \renewcommand{\thesection}{S\arabic{section}}
  \renewcommand{\theequation}{S\arabic{equation}}
  \renewcommand{\thefigure}{S\arabic{figure}}
  \renewcommand{\theHsection}{S\arabic{section}}
  \renewcommand{\thesubsection}{\Alph{subsection}}
  \renewcommand{\theHsubsection}{S\arabic{section}.\Alph{subsection}}
  \renewcommand{\theHfigure}{S\arabic{figure}}
  \renewcommand{\theHequation}{S\arabic{equation}}
  \renewcommand{\theHtable}{S\arabic{table}}
  \secnumdot
}

\makeatother
\begin{document}

\title{Time-of-flight force sensing below the quantum zero-point fluctuation}

\author{Sotatsu~Otabe}
\affiliation{Department of Physics, The University of Tokyo, Japan}

\author{Mitsuyoshi~Kamba}
\affiliation{Department of Physics, The University of Tokyo, Japan}

\author{Yuto~Kojima}
\affiliation{Department of Physics, The University of Tokyo, Japan}

\author{Kiyotaka~Aikawa}
\email[]{aikawa@phys.s.u-tokyo.ac.jp}
\affiliation{Department of Physics, The University of Tokyo, Japan}

\date{\today}

\begin{abstract}
Sensing weak forces through observing a mechanical motion near or below its quantum zero-point fluctuation has been desired in diverse areas. While mechanical oscillators have played a crucial role in such studies, their application to free-fall-type sensing has been elusive, in particular in the quantum regime. Here, we demonstrate sensing a static force of the order of 10 zeptonewtons with a levitated nanomechanical oscillator below the zero-point fluctuation through the rapid modulation of its confining potential. We prepare a squeezed state with a reduced velocity uncertainty by abruptly decreasing the potential. Subsequently, we detect the exerted static force through time-of-flight measurements, where we release the nanoparticle from the potential and measure the displacement during a free fall. Furthermore, time-of-flight measurements allow us to perform quantum state tomography of the squeezed state, from which we reconstruct its Wigner quasiprobability distribution and evaluate the Fisher information for the position measurement to quantify the achievable force sensitivity of our protocol. Our results demonstrate that modulating the trap stiffness serves as a crucial technique for quantum-limited force sensing and paves the way to utilize a levitated nanoparticle as a promising sensing platform beyond the quantum limit with a capability of quantum state tomography. 
\end{abstract}

\maketitle

Detecting weak forces has long been a cutting-edge topic in a wide variety of fields ranging from surface physics~\cite{BUTT20051} and quantum transducers~\cite{Teufel2009,RevModPhys.86.1391} to fundamental physics~\cite{RevModPhys.81.1827}, where a key challenge is to minimize the influence of various fluctuations. With the possibility of cooling their motions to the quantum ground state~\cite{OConnell2010,Chan2011}, mechanical oscillators with a high quality factor offer an ideal platform for sensing very weak forces by reducing the impact of thermal fluctuations. In such systems, the motional uncertainty is fundamentally limited by the zero-point fluctuation. Beating this limit has been one of the central issues in widespread quantum systems, giving rise to extensive studies on quantum squeezing, a key concept to reduce the quantum fluctuation in one quadrature at the cost of the fluctuation in the other quadrature~\cite{doi:10.1126/science.1104149,RevModPhys.89.035002}. Other approaches to overcome the quantum zero-point limit include electromechanical sensing of mechanical motion in hybrid systems~\cite{Teufel2009,doi:10.1126/science.1094419}.\par

Among various mechanical oscillators, a levitated nanoparticle offers unique opportunities of arbitrarily manipulating or even switching off its trapping potential. In combination with the recent advances in cooling their motional degrees of freedom to the ground state~\cite{doi:10.1126/science.aba3993,Magrini2021,Tebbenjohanns2021,PhysRevResearch.4.033051,Kamba:22}, the flexibility in the potential manipulation implies a new class of studies that have been elusive with clamped oscillators. Varying the trapping potential enables a control over the uncertainties on the phase space, resulting in thermal~\cite{PhysRevLett.117.273601} and quantum squeezing~\cite{doi:10.1126/science.ady4652}. Furthermore, releasing the nanoparticle from the potential enables velocity measurements with the time-of-flight (TOF) free expansion of the wavefunction~\cite{PhysRevLett.131.183602,2yzc-fsm3}, which has been a crucial technique in revealing quantum behaviors of trapped neutral atoms~\cite{RevModPhys.80.885}. TOF measurements have also been proposed as an effective means for quantum state tomography of the motion of a single particle~\cite{PhysRevA.83.013803}, which is a key technique to characterize quantum mechanical behaviors and has recently been demonstrated with single atoms~\cite{Brown2023}, while its implementation with a massive object to reconstruct its Wigner function has yet to be realized.\par

In this study, we realize sensing the static gravitational force during the TOF of a levitated nanoparticle near its motional ground state. To suppress the velocity fluctuation to below the zero-point fluctuation, we apply a rapid potential modulation for squeezing the velocity uncertainty before the release from the trap. By observing the ballistic motion of the released nanoparticle due to the gravitational force, we determine the exerted force and reveal the force sensitivity. The sensitivity with velocity squeezing surpasses the limit imposed by the quantum zero-point fluctuation by $\unit[3.5(2)]{dB}$. For the quantitative evaluation of the measured sensitivity, we perform quantum state tomography based on the TOF expansion and reconstruct the Wigner quasiprobability distribution, which allows us to evaluate the Fisher information for the position measurement and confirm that the corresponding expected sensitivities are in good agreement with the measured values.\par


\begin{figure}
\centering
\includegraphics[width=\hsize] {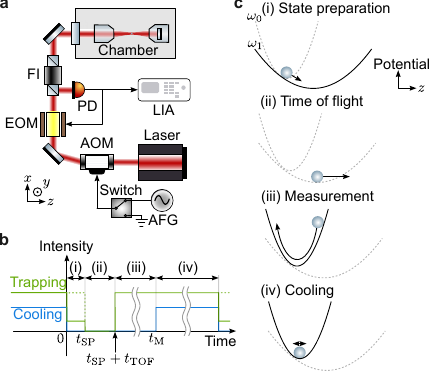}
\caption{\textbf{Outline of TOF force sensing.} 
    \textbf{a}, Schematic of the experimental setup. A laser at a wavelength of $\unit[1551.38]{nm}$ passes through an acousto-optic modulator (AOM) for the rapid intensity control, while an electro-optic modulator (EOM) is used for feedback cooling. The AOM is controlled by an arbitrary function generator (AFG). A single nanoparticle is trapped in an optical lattice formed by retroreflecting the laser with a partially reflective mirror. The back-scattered light is collected through a Faraday isolator (FI) and detected by the PD. A lock-in amplifier (LIA) provides the position signal of the CoM motion along the $z$-axis~\cite{PhysRevLett.131.183602}.
    \textbf{b}, Temporal sequence of the laser intensity for squeezing and TOF measurements.
    \textbf{c}, Schematic of the $z$-axis CoM motion. Solid lines indicate the optical potentials in each phase.
}
\label{fig:setup}
\end{figure}

Our experimental platform is a charge-neutral nanoparticle with a mass of $m=\unit[3.2(2)\times10^{-17}]{kg}$ and the radius of $\unit[150(2)]{nm}$, levitated in an optical lattice (Fig.\,\figref{fig:setup}{a}). The three-dimensional center-of-mass (CoM) motion is continuously monitored via scattered light detected by a photodetector (PD), and the CoM modes along the $x$, $y$, and $z$-axes are cooled by optical feedback~\cite{Kamba:22}. We focus on the $z$-axis CoM mode, which has an oscillation frequency of $\omega_0/2\pi=\unit[221(1)]{kHz}$ and the phonon occupation of $n=0.75(4)$. The gravity is nearly perpendicular to the $z$ direction. To suppress couplings between librational and CoM motions, we spin the nanoparticle at several GHz about the $z$-axis by introducing a circularly polarized component into the trapping light~\cite{doi:10.1126/science.ady4652,PhysRevLett.121.033602,PhysRevLett.121.033603}. The trap region is maintained at a pressure of $\unit[2\times10^{-6}]{Pa}$.\par

The sensing protocol comprises four phases defined by sub-microsecond changes in the trapping and cooling light intensities (Fig.\,\figref{fig:setup}{b}). In phase (i), we prepare a squeezed state by decreasing the oscillation frequency from $\omega_0$ to $\omega_1$ (corresponding to switching the harmonic potential from the dashed to the solid curve in Fig.\,\figref{fig:setup}{c}-(i)). For an abrupt frequency change, the zero-point variances of position and velocity rescale by factors $\omega_0/\omega_1$ and $\omega_1/\omega_0$, respectively. In the rest of this study, instead of the momentum, we regard the velocity directly obtained from the TOF measurements as the conjugate variable to the position. The generated position-squeezed state evolves under the new potential for a time $t_\mathrm{SP}$. The position and velocity variances are interchanged after approximately a quarter period, $T_1/4$, with $T_1=2\pi/\omega_1$.\par

In phase (ii), we release the nanoparticle from the trap and let it freely fly for a duration of $t_\mathrm{TOF}$. Due to the slight tilt of the $z$ axis with respect to the horizontal plane, the nanoparticle undergoes the projection of the gravitational force along the $z$-axis during a free expansion. In phase (iii), by restoring the oscillation frequency to $\omega_0$, we recapture the nanoparticle and measure the absolute value of its position $|z_\mathrm{meas}|$ as an oscillation amplitude. For a sufficiently long TOF duration, this readout provides the velocity before the TOF~\cite{PhysRevLett.131.183602}, which we express as $|v_\mathrm{meas}|=|z_\mathrm{meas}|/t_\mathrm{TOF}$. The cooling beam remains off throughout phases (i)--(iii), for a total measurement window $t_\mathrm{M}=\unit[4]{ms}$. Finally, in phase (iv), the nanoparticle is optically recooled close to the ground state. By repeating such a four-phase sequence approximately $300$ times for the same nanoparticle, we derive the mean value and its fluctuation of the displacement during the TOF.\par

The choice of $\omega_1$ requires a careful consideration on two effects: while lowering $\omega_1$ reduces the velocity uncertainty and enhances the sensitivity, it also broadens the position uncertainty after the state preparation~\cite{kamba2026levitatednanoaccelerometersensitizedquantum}. In this study, we choose $\omega_1/2\pi=\unit[37.3(2)]{kHz}$ such that the broadening of the position variance after the state preparation is negligible in the TOF measurements with sufficiently long $t_\mathrm{TOF}$. In this regime, the position uncertainty has little effect on the sensitivity.\par 


\begin{figure}
\centering
\includegraphics[width=\hsize] {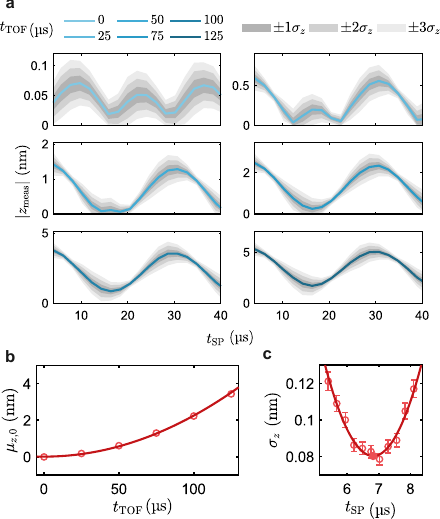}
\caption{\textbf{Dynamics of the nanoparticle during the force sensing protocol.}
    \textbf{a}, Oscillatory dynamics during the force sensing protocol for various values of $t_\mathrm{TOF}$. Solid lines show the estimated mean $\mu_z$, while the shaded regions indicate the estimated ranges of $\pm1$, $\pm2$, and $\pm3$ standard deviations, $\sigma_z$, around the mean. 
    \textbf{b}, Time evolution of the offset extracted from the oscillatory dynamics as a function of $t_\mathrm{TOF}$. The solid line shows a fit with a quadratic function. 
    \textbf{c}, Optimum state-preparation time $t_\mathrm{SP}$ that minimizes $\sigma_z$ for force sensing with $t_\mathrm{TOF}={}$\unit[100]{\textmu s}. Error bars show the standard error from Gaussian fits. The solid line shows a quadratic fit expected near the minimum of $\sigma_z$. The filled circle marks the quadratic vertex at $t_\mathrm{SP}={}$\unit[6.8]{\textmu s}.
}
\label{fig:oscillation}
\end{figure}

\begin{figure}
\centering
\includegraphics[width=\hsize] {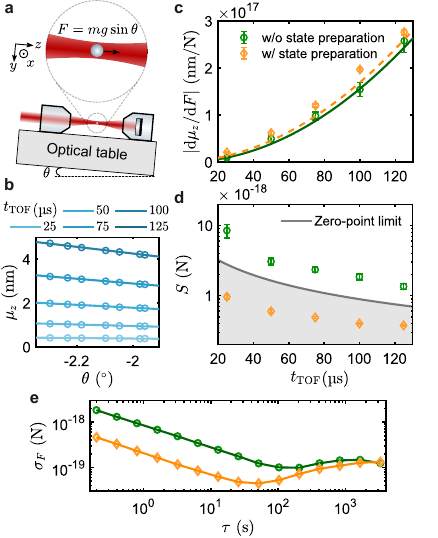}
\caption{ 
    \textbf{Gravitational force sensing below the quantum zero-point fluctuation.}
    \textbf{a}, Schematic of the static force measurement. The optical table and optical lattice are drawn parallel for clarity.
    \textbf{b}, $\mu_z$ achieved with the state preparation for several $\theta$ and $t_\mathrm{TOF}$. 
    \textbf{c}, Susceptibility to the force obtained from linear fits to the measured $\mu_z$ without and with state preparation. Error bars denote the standard error. Solid (dashed) line shows the theoretical prediction without (with) state preparation; these curves are not fit to the data. 
    \textbf{d}, Sensitivity of force sensing as a function of the TOF. Error bars combine the standard error of $\sigma_z$ with the fit uncertainty in \textbf{c}. 
    \textbf{e}, Allan deviation of the force measurements for $t_\mathrm{TOF}={}$\unit[100]{\textmu s} as a function of the integration time $\tau$. Circle and diamond symbols in \textbf{c}, \textbf{d}, and \textbf{e} indicate measurements without and with the state preparation, respectively. 
}
\label{fig:tilt}
\end{figure}

We determine the gravitational force exerted primarily during the TOF from the oscillation amplitudes after recapturing the nanoparticle. From a Gaussian fit to the acquired distribution of $|z_\mathrm{meas}|$, we estimate the mean position $\mu_z$ and its standard deviation $\sigma_z$ (Fig.\,\figref{fig:oscillation}{a}). At longer $t_\mathrm{TOF}$, the contribution of the position uncertainty after the state preparation is reduced and $\mu_z$ is determined more precisely. The oscillation amplitude of $\mu_z$ depends on the gravitational force and an intensity-dependent phase shift of the trapping laser (Supplementary Information). To extract the force exerted during the TOF, we obtain the offset values $\mu_{z,0}$ from fitting $\mu_z$ with the absolute value of a sinusoidal function. We find that $\mu_{z,0}$ increases quadratically with $t_\mathrm{TOF}$ (Fig.\,\figref{fig:oscillation}{b}), which is consistent with a ballistic motion in the presence of a static force along the $z$-axis, $F=mg\sin\theta$ (Fig.\,\figref{fig:tilt}{a}), where $g$ is the gravitational acceleration and $\theta$ is the small tilt of the optical lattice with respect to the horizontal plane. Using the known value of $g=9.798$ m/$\mathrm{s}^2$ in Tokyo, we find $\theta = -2.62(5)^\circ$. For force sensing with the state preparation, $t_\mathrm{SP}$ is carefully chosen to minimize $\sigma_z$ (Fig.\,\figref{fig:oscillation}{c}), while we also study the performance of force sensing without the state preparation by taking $t_\mathrm{SP}=0$~\textmu s for comparison. \par

We reveal the susceptibility in force sensing by measuring the variation of the gravitational force projected onto the $z$-axis with respect to the tilt of the optical table. From the linear fits on the measured $\mu_z$ for the various tilt angles $\theta$, we derive the susceptibility to the force $\mathrm{d}\mu_z/\mathrm{d}F$ (Fig.\,\figref{fig:tilt}{b,c}). Such measurements are performed for various values of $t_\mathrm{TOF}$. For the free expansion during TOF, the gravity-induced displacement and the susceptibility with small $\theta$ scale as $t_\mathrm{TOF}^2$, regardless of whether the state preparation is applied. Assuming a ballistic motion, the theoretical susceptibility without the state preparation can be calculated as $ t_\mathrm{TOF}^2/2m =\unit[1.6\times 10^{17}]{nm/N}$ for $t_\mathrm{TOF}={}$\unit[100]{\textmu s}, whereas the state preparation slightly increases the susceptibility to $\unit[1.7 \times 10^{17}]{nm/N}$. The increased susceptibility is understood as a result of the gravity-induced motion introduced by ramping down the potential for the state preparation, with an oscillation period of $T_1$ and an amplitude depending on the projection of the gravity along the $z$-axis (Supplementary Information).\par

We define the sensitivity in force sensing as $S=\sigma_z/|\mathrm{d}\mu_z/\mathrm{d}F |$, where $\sigma_z$ is the measured position fluctuation immediately after the TOF. A comparison of the sensitivity observed with and without the state preparation is shown in Fig.\,\figref{fig:tilt}{d}. With the state preparation, the sensitivity improves markedly owing to the narrowed velocity variance below the zero-point limit, which is calculated for $n=0$ using the theoretical susceptibility of the state-prepared protocol. When background-gas heating is negligible, $\sigma_z$ increases linearly with $t_\mathrm{TOF}$, implying that a longer TOF improves the sensitivity. In the current setup, the TOF is practically limited to around \unit[200]{\textmu s} due to the vertical displacement, which results in the particle loss at a longer TOF. Possible solutions to this issue are to place another trap for recapturing~\cite{47pz-74fc} or to use a vertical optical lattice~\cite{Stickler_2018}. Moreover, increasing the squeezing level by further reducing the intensity for the state preparation is expected to enhance the sensitivity.\par

We also investigate the long-term stability of force sensing by recording the data for about $\unit[10^4]{s}$, from which we derive the Allan deviation for force sensing (Fig.\,\figref{fig:tilt}{e}). We find that the Allan deviation for short integration times $\tau$ follows $S/\sqrt{f_{s} \tau}$, with $f_s = \unit[5]{Hz}$ being the sampling frequency, and decreases down to about $\unit[4\times 10^{-20}]{N}$ at an integration time of $\unit[50]{s}$. At a longer integration time, the Allan deviation increases, suggesting that the long-term stability is limited by the thermal fluctuations of the experimental setup. The observed sensitivity is comparable to that of a recently reported sensing scheme based on ramping down the optical potential without TOF measurements~\cite{kamba2026levitatednanoaccelerometersensitizedquantum}.\par


\begin{figure}
\centering
\includegraphics[width=\hsize] {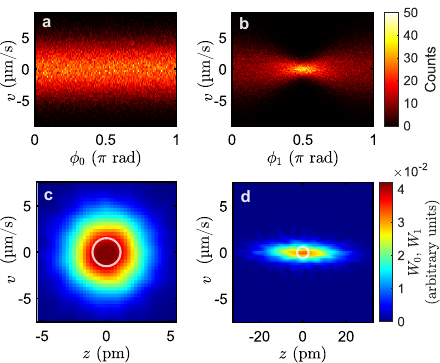}
\caption{\textbf{Quantum state tomography of a levitated nanoparticle.}
    \textbf{a},\textbf{b}, Velocity sinograms without and with the state preparation for $t_\mathrm{TOF}={}$\unit[100]{\textmu s}. The tomography phase is sampled in $300$ steps from $0$ to about $\unit[\pi]{rad}$. At each phase, we acquire approximately 600 traces. For the sinogram construction, each phase-resolved velocity histogram is centered by subtracting its mean offset before the tomographic reconstruction.
    \textbf{c},\textbf{d}, Reconstructed Wigner functions of the motional states at $t_\mathrm{SP}={}$\unit[0]{\textmu s} and $t_\mathrm{SP}={}$\unit[6.8]{\textmu s}, respectively. For clearly visualizing the narrowed velocity uncertainty after the state preparation, the vertical range is chosen to be common to both panels. The zero-point fluctuation contours for the initial potential and for the potential during state preparation are indicated as solid circles in \textbf{c} and \textbf{d}, respectively.
}
\label{fig:qst}
\end{figure}

To gain further insight into the performance of our measurements,  we perform quantum state tomography based on the TOF free expansion of the wavefunction~\cite{PhysRevA.83.013803} and derive the Fisher information for the force exerted on the nanoparticle. We evaluate the velocity statistics relevant to the state preparation by analyzing the phase-resolved velocity projections. Figure\,\figref{fig:qst}{a,b} shows velocity-projection sinograms without and with the state preparation, respectively. The sinogram without the state preparation is obtained by switching the trapping power as indicated by dashed line in Fig.\,\figref{fig:setup}{b}. Here $t_\mathrm{SP}$ marks the TOF start time and is converted to the phase $\phi_0=2\pi t_\mathrm{SP}/T_0$, where $T_0=2\pi/\omega_0$. In this case, the velocity variance is observed to be independent of $\phi_0$. In contrast, with the state preparation, we convert $t_\mathrm{SP}$ to the phase $\phi_1=2\pi t_\mathrm{SP}/T_1$. The velocity variance oscillates at $2\omega_1$, consistent with the generation of a motional squeezed state~\cite{doi:10.1126/science.ady4652}. \par

Phase-resolved measurements with a sufficient angular resolution allow us to reconstruct the quasiprobability distribution of the nanoparticle motion. Figure\,\figref{fig:qst}{c,d} shows the motional Wigner functions reconstructed by the maximum likelihood estimation (Appendix), without and with the state preparation, $W_0$ and $W_1$, respectively. Without state preparation, the motional state is close to the ground state, and the estimated root-mean-square velocity is $1.89(1)\times v_0^\mathrm{zpf}$, where $v_0^\mathrm{zpf}=\sqrt{\hbar \omega_0/2m}$ is the zero-point fluctuation of the velocity in the initial trapping potential. With the state preparation, the velocity variance is reduced by choosing an appropriate state preparation time; the minimum estimated root-mean-square velocity is $0.520(3) \times v_0^\mathrm{zpf}$, realizing a sensitivity below the zero-point fluctuation at the oscillation frequency $\omega_0$.\par

Using the density matrices reconstructed by quantum state tomography, we calculate the Fisher information for the position measurement (Supplementary Information). This corresponds to $1/S^2$ and gives the expected sensitivities at $t_\mathrm{TOF}={}$\unit[100]{\textmu s} as $\unit[1.8(1)\times 10^{-18}]{N}$ and $\unit[4.2(3)\times 10^{-19}]{N}$ for the cases without and with the state preparation, respectively. These values are in reasonable agreement with the measured sensitivities in Fig.~\figref{fig:tilt}{d}, which are $\unit[1.9(2)\times 10^{-18}]{N}$ and $\unit[4.01(9)\times 10^{-19}]{N}$. \par


In conclusion, we demonstrate force sensing below the quantum zero-point fluctuation in the original trapping potential through TOF measurements of a levitated nanoparticle. We show that the state preparation with the abrupt potential modulation squeezes the velocity uncertainty and achieves a sensitivity beyond the value limited by the zero-point fluctuation. Furthermore, the TOF measurements enable us to realize quantum state tomography of the motional state and reconstruct its Wigner function. This also allows us to estimate the Fisher information for the position measurement, yielding expected sensitivities in reasonable agreement with the measured values. More generally, the quantum Fisher information sets the ultimate precision bound for a given quantum state through the quantum Cram\'{e}r-Rao bound~\cite{kamba2026levitatednanoaccelerometersensitizedquantum,Wiseman_Milburn_2009}.\par

In a broader context, related ideas based on stiffness modulation have also been explored in cavity optomechanics, where parametric modulation with an optical spring was used to generate a thermal squeezed state of a microgram-scale mechanical oscillator~\cite{PhysRevLett.112.023601}. More recently, optical-spring engineering has enabled sensitivities surpassing the standard quantum limit in the free-mass picture~\cite{PhysRevLett.133.113602}. Extending the rapid trap-stiffness modulation demonstrated here to more massive oscillators through optical spring may open a route toward the verification of nonclassical states and the realization of quantum-enhanced sensing with larger mechanical systems.\par

Quantum state tomography demonstrated here is an important tool for exploring the quantum mechanical behaviors of levitated nanoparticles. Efforts to generate exotic mechanical quantum states based on non-Gaussian states are accelerating~\cite{doi:10.1073/pnas.2306953121,PhysRevA.110.033511,PhysRevResearch.7.013171}, underscoring motional tomography as a key tool for validating and quantifying these resources~\cite{RashidTorosUlbricht+2017+17+25}. The gravity sensing with the tomographic validation will open the prospects for testing the gravity-related modifications of quantum dynamics, including gravitational decoherence models~\cite{PhysRevA.40.1165,Penrose1996,Pikovski2015,Kaltenbaek_2023} and Schr\"{o}dinger--Newton-type semiclassical self-gravity effects~\cite{PhysRevLett.110.170401,PhysRevD.93.096003,PhysRevA.93.062102,altamura2025enhancementeffectsschrodingernewtonequation}.\par

\textit{Note added}: While finalizing this manuscript, we became aware of a related preprint reporting coherent mechanical amplification of impulsive forces below the zero-point momentum fluctuation of an optically levitated nanoparticle~\cite{skrabulis2026nanomechanicalsensorresolvingimpulsive}, complementary to our demonstration of TOF sensing of a static force below the zero-point velocity fluctuation and motional quantum-state tomography.

\beginmethods

\subsection{Experimental setup}

A single-frequency laser at a wavelength of $\unit[1551.38]{nm}$ and with a power of $\unit[185]{mW}$ is focused with an objective lens (NA$=0.85$) and approximately a quarter of the incident power is retro-reflected to form an optical lattice. We load nanoparticles by blowing silica powder placed near the trapping region with a pulsed laser at $\unit[532]{nm}$ at pressures of approximately $\unit[5\times10^2]{Pa}$. At around $\unit[3\times10^2]{Pa}$, we apply a positive high voltage to induce a corona discharge and provide a positive charge on the nanoparticle. Then we evacuate the chamber with optical feedback cooling for the translational motions. We neutralize the nanoparticle with an ultraviolet light at around $\unit[2 \times 10^{-5}]{Pa}$. At around $\unit[1 \times 10^{-5}]{Pa}$, we turn off the turbo-molecular pump and the scroll pump, and continue evacuation using an ion pump and a titanium sublimation pump.

\subsection{Characterization of the trapped nanoparticle}

We estimate the density and the radius of the trapped nanoparticle via the two independent measurements. First, we measure the heating rate at around $\unit[4.94]{Pa}$, which is given by the background gas collisions~\cite{PhysRevLett.109.103603,Vovrosh:17,PhysRevA.99.051401}. We measure the pressure with an accuracy of $\unit[0.5]{\%}$ via a capacitance gauge. Second, we measure the heating rate at around $\unit[2.0 \times 10^{-6}]{Pa}$, which is determined dominantly by photon recoil heating and is more sensitive to the radius than the heating rate at $\unit[4.94]{Pa}$. By combining these results, we determine the radius and the density of the nanoparticle to be $\unit[150(2)]{nm}$ and $\unit[2.29(3) \times 10^3]{kg/m^3}$, respectively. The mass of the nanoparticle is $m=3.2(2) \times \unit[10^{-17}]{kg}$.\par

The CoM temperatures are obtained by comparing the areas of the PSDs with and without cooling, as has been performed in previous studies~\cite{PhysRevLett.109.103603,Vovrosh:17,PhysRevA.99.051401}. To avoid the influence of the increase in the internal temperature of nanoparticles at high vacuum due to laser absorption~\cite{PhysRevA.97.043803}, we take the uncooled data at around $\unit[5]{Pa}$. The PSD is averaged over 1000 times to minimize thermal fluctuations. In addition, we take more than 10 averaged traces to derive the mean value of the area of the PSD as well as its standard deviation. We find that the typical thermal fluctuation of the area of the PSD is lower than $\unit[5]{\%}$ for both cooled and uncooled data. Thus, we estimate the systematic error in determining the temperatures of CoM motions to be about $\unit[7]{\%}$. 

\subsection{Calibration of the position signal with a RF frequency shift}

We rely on the RF calibration procedure to determine the absolute magnitude of the oscillation signal~\cite{doi:10.1126/science.ady4652}. The AOM for varying the laser intensity and turning off the laser is driven by an RF switch (ZASWA-2-50DRA+, Mini-Circuits). While the RF frequency is kept constant at $\unit[80]{MHz}$ for trapping, we induce an oscillation by shifting the RF frequency by $\unit[1.4]{MHz}$ for the calibration with the shift of the laser frequency. We obtain the calibration factor between the position and the voltage from the amplitude of the oscillation. The device for recording the oscillation is the same as that for TOF measurements. We take about 300 traces for each time duration, from which the center of the distribution is extracted as a mean oscillation amplitude. 

\subsection{Data acquisition and analysis for velocity distributions}

To measure the velocity of the nanoparticle after free expansion, we record the photodetector signal after passing through two fourth-order high-pass filters (cutoff frequencies of $\unit[105]{kHz}$ and $\unit[150]{kHz}$) and an eighth-order high-pass filter (cutoff frequency of 160~kHz) with a lock-in amplifier (MFLI, Zurich Instruments). These filters are required to remove low-frequency noise.\par

We perform the following data processing for each single measurement in order to further suppress low frequency noises from the recorded data. The data is filtered through a 10th order finite-impulse-response filter centered at $\unit[221]{kHz}$ with a bandwidth of $\unit[20]{kHz}$. The amplitude is extracted from the envelope of the filtered trace. By dividing the amplitude by $t_{\rm TOF}$, we obtain the velocity of the nanoparticle for each measurement. For the force-sensitivity measurements, we repeat the sequence approximately $300$ times under each experimental condition and construct a distribution of the measured displacement.
For the tomography measurements, the tomography phase is sampled in $300$ steps from $0$ to roughly $\pi$ rad, and approximately $600$ traces are acquired at each phase to construct a velocity histogram for the sinogram.

\subsection{Summary of quantum state tomography}

Maximum likelihood estimation (MLE) is a common statistical method used to estimate the parameters of a probability distribution. Assuming a certain probability model for the measured data, we can formulate the likelihood function and maximize it to obtain the best estimate of the unknown parameters. In the case of quantum state tomography, one can evaluate the likelihood of the measurement results given a particular quantum state as a density matrix in some basis. Thus, maximizing the likelihood allows us to find the density matrix most likely to the observed data by optimizing the complex-valued matrix elements. This method has an advantage that the reconstructed density matrix is always physical, meaning it is positive semidefinite and has a trace of one. In contrast, other methods such as the inverse Radon transform may yield unphysical density matrices due to statistical fluctuations in the data. Therefore, MLE is widely used for quantum state reconstruction in various quantum systems. In this study, we use a diluted iterative MLE algorithm for homodyne measurements~\cite{AILvovsky_2004, PhysRevA.75.042108}.

\subsection{Calculation of the Fisher information}

To quantify the achievable sensitivity of our protocol, we use the density matrices reconstructed by quantum state tomography as input states and evaluate the classical Fisher information associated with the position measurement after TOF expansion for estimation of the parameter $\theta$. We treat both cases, with and without state preparation. The full derivation is given in the Supplementary Information.

\subsection{}

\textbf{Data Availability} All data that support the findings of the study are provided in the main text and in the Supplementary Information.\par

\textbf{Code Availability} The conclusions of this study do not depend on code or algorithms beyond standard numerical evaluations.


%

\textbf{Acknowledgements} We thank M. Kozuma, T. Mukaiyama, K. Funo, K. Kumasaki, I. Layton, T. Sagawa, S. Sugiura, K. Shiraishi, and M. Nakagawa for
fruitful discussions. This work is supported by the Murata Science Foundation, the Mitsubishi Foundation, the Challenging Research Award, the 'Planting Seeds for Research' program, Yoshinori Ohsumi Fund for Fundamental Research, and STAR Grant funded by the Tokyo Tech Fund, Research Foundation for Opto-Science and Technology, JSPS KAKENHI (Grants No. JP16K13857, JP16H06016, JP19H01822, and JP22K18688), JST PRESTO (Grant No. JPMJPR1661), JST ERATO (Grant No.JPMJER2302), JST CREST (Grant No. JPMJCR23I1), and JST COI-NEXT (Grant No. JPMJPF2015).\par
 
\textbf{Author contributions} K.\,A., M.\,K., and S.\,O. designed the experiments. S.\,O., M.\,K., and Y.\,K., performed measurements and analyzed the data. Y.\,K. performed calculations on the reconstruction of Wigner functions from the experimental data. M.\,K. performed calculations on the Fisher information. All authors discussed the results and contributed to writing the manuscript.\par

\textbf{Competing Interests} The authors declare no competing interests.\par

\beginsuppinfo

\section{Derivation of sensitivity metrics}

To analyze the presented protocol, we describe the CoM motion of the nanoparticle along the z-axis within a one-dimensional phase-space formalism under a constant force $mg\sin\theta$. We consider a nanoparticle with initial position $z(0)$, momentum $p(0)$, and covariance matrix $V(0)=\mathrm{diag} (\kappa\hbar/2m\omega_0, \kappa\hbar m\omega_0/2)$, with $\kappa=2n+1$. Position and momentum after state preparation can be written as
\begin{align}
    \begin{pmatrix}
        z(t_\mathrm{SP})\\p(t_\mathrm{SP})
    \end{pmatrix}
    =&
    \begin{pmatrix}
        \cos\omega_1 t_\mathrm{SP} & \frac{1}{m\omega_1}\sin\omega_1 t_\mathrm{SP}\\
        -m\omega_1\sin\omega_1 t_\mathrm{SP} & \cos\omega_1 t_\mathrm{SP}
    \end{pmatrix}
    \begin{pmatrix}
        z(0)\\p(0)
    \end{pmatrix}\nonumber\\
    &+
    \begin{pmatrix}
        \frac{g\sin\theta}{\omega^2_1}[1-\cos\omega_1 t_\mathrm{SP}]\\
        \frac{mg\sin\theta}{\omega_1}\sin\omega_1t_\mathrm{SP}
    \end{pmatrix}\nonumber\\
    =&A_1(t_\mathrm{SP})\begin{pmatrix}
        z(0)\\p(0)
    \end{pmatrix}+\bm{b}_1(t_\mathrm{SP}).
\end{align}
Covariance matrix after state preparation can be calculated as
\begin{align}
    V^\mathrm{SP}=A_1(t_\mathrm{SP})V(0)A_1(t_\mathrm{SP})^\top=\begin{pmatrix}
         V_{zz}^\mathrm{SP} & V_{zp}^\mathrm{SP}\\
        V_{zp}^\mathrm{SP} & V_{pp}^\mathrm{SP}
    \end{pmatrix},
\end{align}
with
\begin{align}
    V_{zz}^\mathrm{SP}=&\frac{\kappa\hbar}{2m\omega_0}\left[\cos^2\omega_1 t_\mathrm{SP}+\left(\frac{\omega_0}{\omega_1}\right)^2\sin^2\omega_1 t_\mathrm{SP}\right],\\
    V_{zp}^\mathrm{SP}=&\frac{\kappa\hbar}{4}\left(\frac{\omega_0}{\omega_1}-\frac{\omega_1}{\omega_0}\right)\sin2\omega_1 t_\mathrm{SP},\\
    V_{pp}^\mathrm{SP}=&\frac{\kappa\hbar m\omega_0}{2}\left[\cos^2\omega_1 t_\mathrm{SP}+\left(\frac{\omega_1}{\omega_0}\right)^2\sin^2\omega_1 t_\mathrm{SP}\right].
\end{align}
Position and momentum after TOF expansion can be written as
\begin{align}
    &\begin{pmatrix}
        z(t_\mathrm{SP}+t_\mathrm{TOF})\\
        p(t_\mathrm{SP}+t_\mathrm{TOF})
    \end{pmatrix}\nonumber\\
    =&\begin{pmatrix}
        1 & \frac{t_\mathrm{TOF}}{m}\\
        0 & 1
    \end{pmatrix}\begin{pmatrix}
        z(t_\mathrm{SP})\\
        p(t_\mathrm{SP})
    \end{pmatrix}+\begin{pmatrix}
        \frac{1}{2}g\sin\theta t_\mathrm{TOF}^2\\
        mg\sin\theta t_\mathrm{TOF}
    \end{pmatrix}\nonumber\\
    =&A_2(t_\mathrm{TOF})\begin{pmatrix}
        z(t_\mathrm{SP})\\
        p(t_\mathrm{SP})
    \end{pmatrix}+\bm{b}_2(t_\mathrm{TOF}).\label{eq:zafterTOF}
\end{align}
Covariance matrix after TOF expansion can be calculated as
\begin{align}
    V^\mathrm{TOF}=A_2(t_\mathrm{TOF})V^\mathrm{SP}A_2(t_\mathrm{TOF})^\top=\begin{pmatrix}
         V_{zz}^\mathrm{TOF} & V_{zp}^\mathrm{TOF}\\
        V_{zp}^\mathrm{TOF} & V_{pp}^\mathrm{TOF}
    \end{pmatrix},
\end{align}
with
\begin{align}
    V_{zz}^\mathrm{TOF}=&V_{zz}^\mathrm{SP}+\frac{2t_\mathrm{TOF}}{m}V_{zp}^\mathrm{SP}+\left(\frac{t_\mathrm{TOF}}{m}\right)^2V_{pp}^\mathrm{SP},\\
    V_{zp}^\mathrm{TOF}=&V_{zp}^\mathrm{SP}+\frac{t_\mathrm{TOF}}{m}V_{pp}^\mathrm{SP},\\
    V_{pp}^\mathrm{TOF}=&V_{pp}^\mathrm{SP}.
\end{align}
\par

Without state preparation ($t_\mathrm{SP}=0$), particle position and position variance after TOF expansion can be calculated as
\begin{align}
    \left.z(t_\mathrm{SP}+t_\mathrm{TOF})\right|_{t_\mathrm{SP}=0}=&z(0)+\frac{t_\mathrm{TOF}}{m}p(0)+\frac{1}{2}g\sin\theta t_\mathrm{TOF}^2,\\
    \left.V_{zz}^\mathrm{TOF}\right|_{t_\mathrm{SP}=0}=&\frac{\kappa\hbar}{2m\omega_0}\left[1+(\omega_0t_\mathrm{TOF})^2\right].
\end{align}
With state preparation, sufficiently long TOF eliminates the effect of the position variance after state preparation $V^\mathrm{SP}_{zz}$. The momentum variance after state preparation $V_{pp}^\mathrm{SP}$ is minimized at $t_\mathrm{SP}=\pi/2\omega_1$, we obtain
\begin{align}
    \left.z(t_\mathrm{SP}+t_\mathrm{TOF})\right|&_{t_\mathrm{SP}=\pi/2\omega_1}
    =\frac{p(0)}{m\omega_1}-\omega_1 t_\mathrm{TOF}z(0)\nonumber\\
    &+g\sin\theta\left[\frac{1}{\omega_1^2}+\frac{t_\mathrm{TOF}}{\omega_1}+\frac{t_\mathrm{TOF}^2}{2}\right],\\
    \left.V_{zz}^\mathrm{TOF}\right|_{t_\mathrm{SP}=\pi/2\omega_1}=&\frac{\kappa\hbar}{2m\omega_0}\left[\left(\frac{\omega_0}{\omega_1}\right)^2+\left(\omega_1t_\mathrm{TOF}\right)^2\right].\label{eq:VzzTOFwithSP}
\end{align}
In the small-angle approximation ($\theta \ll 1$), the force sensitivity can be written as $S\simeq mg\sqrt{V_{zz}^\mathrm{TOF}}/|\mathrm{d}z_\mathrm{meas}/\mathrm{d}\theta|$, where $z_\mathrm{meas}=z(t_\mathrm{SP}+t_\mathrm{TOF})-g\sin\theta/\omega_0^2$ is the position measured relative to the gravity-shifted origin. Substituting $z(0)=g\sin\theta/\omega_0^2$ and $p(0)=0$, we obtain
\begin{align}
    \left.z_\mathrm{meas}\right|_{t_\mathrm{SP}=0}=&\frac{1}{2}g\sin\theta t_\mathrm{TOF}^2,\\
    \left.z_\mathrm{meas}\right|_{t_\mathrm{SP}=\pi/2\omega_1}=&g\sin\theta\left[(1+\omega_1 t_\mathrm{TOF})\left(\frac{1}{\omega_1^2}-\frac{1}{\omega_0^2}\right)\right.\nonumber\\
    &\left.+\frac{1}{2}t_\mathrm{TOF}^2\right].
\end{align}
Without the state preparation, the sensitivity has a lower bound determined by the possible TOF duration and cannot beat this limit even if $n=0$:
\begin{equation}
    S^\mathrm{TOF}\geq \frac{\sqrt{2m\hbar\omega_0}}{t_\mathrm{TOF}}.
\end{equation}
For a sufficiently long TOF duration ($t_\mathrm{TOF} \gg 1/\omega_1$), state preparation enables force sensitivity below the zero-point limit, with $S$ scaled by a factor of $\omega_1/\omega_0$.\par

\section{Deterioration of the position variance due to background gas heating}

This section derives the impact of background gas heating on the covariance matrix. The dynamics of the nanoparticle are described by the quantum Langevin equation~\cite{gardiner2004quantum}:
\begin{align}
    \frac{\mathrm{d}z}{\mathrm{d}t}=&\frac{p}{m},\nonumber\\
    \frac{\mathrm{d}p}{\mathrm{d}t}=&-m\omega^2(t)[z-z_k(t)]+ma-\gamma p+\xi(t),
\end{align}
where $\omega(t)$ is the resonance frequency, $z_k(t)$ is the potential shift proportional to the light intensity~\cite{kamba2026levitatednanoaccelerometersensitizedquantum}, and $\gamma$ and $\xi(t)$ are the damping rate and stochastic thermal force due to the background gas, respectively. According to the fluctuation-dissipation theorem~\cite{PhysRev.83.34}, the autocorrelation of this thermal force is given by $\langle\xi(t)\xi(0)\rangle=2m\gamma k_\mathrm{B}T_0 \delta(t)$, where $T_0$ is the temperature of the background gas. Assuming a sufficiently short timescale where the background gas damping is negligible, background gas heating adds the following term to the time derivative of momentum variance:
\begin{equation}
    \frac{\mathrm{d}}{\mathrm{d}t}V_{pp}^\mathrm{BG}=2mk_\mathrm{B}\Gamma_\mathrm{BG},
\end{equation}
where $\Gamma_\mathrm{BG}=\gamma T_0$ is the heating rate due to the background gas. All other elements of matrix $\mathrm{d}V^\mathrm{BG}/\mathrm{d}t$ are zero: $\mathrm{d}V_{zz}^\mathrm{BG}/\mathrm{d}t=\mathrm{d}V_{zp}^\mathrm{BG}/\mathrm{d}t=\mathrm{d}V_{pz}^\mathrm{BG}/\mathrm{d}t=0$. Background gas heating during the state preparation adds the following component to the covariance matrix:
\begin{align}
    V^\mathrm{SP-BG}&=\int_0^{t_\mathrm{SP}}A_1(t_\mathrm{SP}-t)\frac{\mathrm{d}V^\mathrm{BG}}{\mathrm{d}t}A_1(t_\mathrm{SP}-t)^\top\mathrm{d}t\nonumber\\
    &=\begin{pmatrix}
         V_{zz}^\mathrm{SP-BG} & V_{zp}^\mathrm{SP-BG}\\
        V_{zp}^\mathrm{SP-BG} & V_{pp}^\mathrm{SP-BG}
    \end{pmatrix},
\end{align}
with
\begin{align}
    V_{zz}^\mathrm{SP-BG}=&\frac{k_\mathrm{B}\Gamma_\mathrm{BG}}{m\omega_1^2}\left(t_\mathrm{SP}-\frac{\sin2\omega_1 t_\mathrm{SP}}{2\omega_1}\right),\\
    V_{zp}^\mathrm{SP-BG}=&\frac{k_\mathrm{B}\Gamma_\mathrm{BG}}{2\omega_1^2}\left(1-\cos2\omega_1 t_\mathrm{SP}\right),\\
    V_{pp}^\mathrm{SP-BG}=&mk_\mathrm{B}\Gamma_\mathrm{BG}\left(t_\mathrm{SP}+\frac{\sin2\omega_1 t_\mathrm{SP}}{2\omega_1}\right).
\end{align}
The covariance matrix after TOF is modified as follows:
\begin{align}
    V^\mathrm{TOF-tot}=&A_2(t_\mathrm{TOF})\left(V^\mathrm{SP}+V^\mathrm{SP-BG}\right)A_2(t_\mathrm{TOF})^\top\nonumber\\
    &+V^\mathrm{TOF-BG},
\end{align}
where $V^\mathrm{TOF-BG}$ is the additional component due to background gas heating during TOF expansion:
\begin{align}
    V^\mathrm{TOF-BG}&=\int_0^{t_\mathrm{TOF}}A_2(t_\mathrm{TOF}-t)\frac{\mathrm{d}V^\mathrm{BG}}{\mathrm{d}t}A_2(t_\mathrm{TOF}-t)^\top\mathrm{d}t\nonumber\\
    &=\begin{pmatrix}
         V_{zz}^\mathrm{TOF-BG} & V_{zp}^\mathrm{TOF-BG}\\
        V_{zp}^\mathrm{TOF-BG} & V_{pp}^\mathrm{TOF-BG}
    \end{pmatrix},
\end{align}
with
\begin{align}
    V_{zz}^\mathrm{TOF-BG}=&\frac{2k_\mathrm{B}\Gamma_\mathrm{BG}}{3m}t_\mathrm{TOF}^3,\label{eq:VzzTOFBG}\\
    V_{zp}^\mathrm{TOF-BG}=&k_\mathrm{B}\Gamma_\mathrm{BG}t_\mathrm{TOF}^2,\\
    V_{pp}^\mathrm{TOF-BG}=&2mk_\mathrm{B}\Gamma_\mathrm{BG}t_\mathrm{TOF}.
\end{align}

\section{Parameter estimation of the oscillatory motion}

\begin{figure}
	\centering
	\includegraphics[width=\hsize]{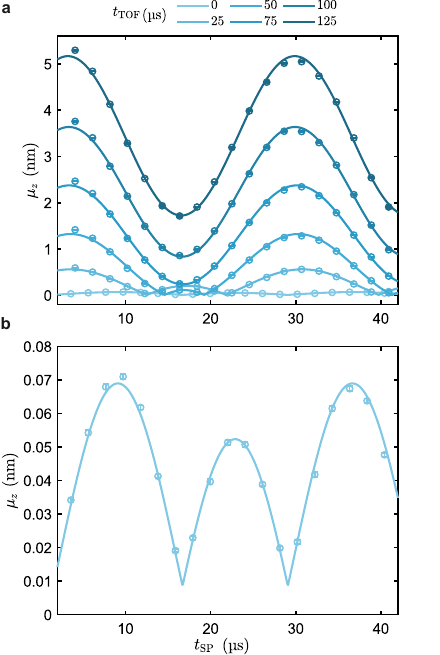}
	\caption{
    \textbf{Position oscillation induced by the state preparation.}
    \textbf{a}, Measured absolute value of the mean particle position $\mu_z$ for several TOF durations $t_\mathrm{TOF}$. Error bars show the standard error from Gaussian fits. Solid lines show theoretical fits with the absolute value of a sinusoidal function with an offset. \textbf{b}, Expanded plot for $t_\mathrm{TOF}={}$\unit[0]{\textmu s}. This oscillation is roughly in quadrature with the momentum measurements with sufficiently long TOF durations because it is proportional to the particle displacement at the end of state preparation.
    } 
	\label{fig:oscillationattTOF0}
\end{figure}

The oscillation induced by the state preparation (Fig.~\figref{fig:oscillationattTOF0}{a}) can be expressed as
\begin{equation}
    z_\mathrm{osci}=|b_1+b_2\cos(b_3 t_\mathrm{SP}+b_4)|+b_5,
\end{equation}
where $b_1$ to $b_5$ are fitting parameters. The measurement offset $b_5$ originates from the measurement noise in MFLI, which is not negligible when $t_\mathrm{TOF}=0$~\textmu s; we estimate this value by fitting the oscillation at $t_\mathrm{TOF}=0$~\textmu s (Fig.\,\figref{fig:oscillationattTOF0}{b}), yielding a value of $b_5=\unit[9(1)]{pm}$. We perform the fitting for $t_\mathrm{TOF}\geq{}$\unit[25]{\textmu s} by fixing $b_5$. We obtain the oscillation offset $b_1=\mu_{z,0}$ from the fitting, which is written as the following function derived from Eq\,\eqref{eq:zafterTOF}:
\begin{equation}
    b_1=\frac{1}{2}g \sin\theta\left(t_\mathrm{TOF}^2-\frac{1}{\omega_0^2}+\frac{1}{\omega_1^2}\right).
\end{equation}
We estimate the tilt of the optical table by fitting with $\theta$ as a free parameter and use the fitted value to calibrate the offset of the inclinometer, whose measurement accuracy is $\unit[3 \times 10^{-4}]{^\circ}$. Here, $\omega_1$ is estimated from the average value of $b_3$ for measurements at various $t_\mathrm{TOF}$. The oscillation amplitude $b_2$ cannot be explained by Eq.\,\eqref{eq:zafterTOF} because it is affected by $z_k(t)$.

\section{Determination of the squeezing parameter}

\begin{figure}
	\centering
	\includegraphics[width=\hsize]{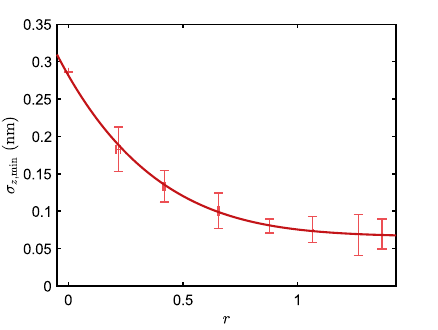}
	\caption{
    \textbf{Minimum value of $\sigma_z$ as a function of the squeezing parameter.} The TOF duration is fixed to $t_\mathrm{TOF}={}$\unit[100]{\textmu s}. Vertical and horizontal error bars indicate the standard errors obtained from the respective estimation procedures.
    }
	\label{fig:rDetermination}
\end{figure}

The squeezing parameter $r$ is determined by the ratio of the trapping powers in cooling and state preparation phases, $I_0$ and $I_1$, respectively, which are proportional to square of $\omega_0$ and $\omega_1$~\cite{doi:10.1126/science.ady4652}:
\begin{equation}
    r=\frac{1}{4}\log\left(\frac{I_0}{I_1}\right)=\frac{1}{2}\log\left(\frac{\omega_0}{\omega_1}\right).
\end{equation}
We estimated the respective $\omega_1$ values from the measured oscillations for various AOM drive amplitudes relative to $t_\mathrm{SP}$. For each estimated value of $r$, we measure $\sigma_z$ near $t_\mathrm{SP}$ that minimizes $\sigma_z$, and estimate the minimum value $\sigma_{z,\mathrm{min}}$ by fitting to a quadratic function. The minimum value of $\sigma_z$ obeys following function obtained from Eq.\,\eqref{eq:VzzTOFwithSP}:
\begin{equation}
    \left.\sigma_z\right|_{t_\mathrm{SP}=\pi/2\omega_1}=\sqrt{\frac{\kappa\hbar}{2m\omega_0}\left[\mathrm{e}^{4r}+(\omega_0t_\mathrm{TOF})^2\mathrm{e}^{-4r}\right]}.\label{eq:sigmazatT1per4min}
\end{equation}
For actual data subject to heating and measurement noise, the following equation describes the experimental results well~\cite{doi:10.1126/science.ady4652}:
\begin{equation}
    \frac{m\sigma_{z,\mathrm{min}}}{t_\mathrm{TOF}\sqrt{\hbar m\omega_0/2}}=\sqrt{V_\mathrm{n}+V_\mathrm{ini}\mathrm{e}^{-4r}},
\end{equation}
where $V_\mathrm{ini}$ is the normalized momentum variance, and $V_\mathrm{n}$ is the contribution from measurement uncertainties including the influence of position variance term. We obtain $V_\mathrm{ini}=3.27(3)$ and $V_\mathrm{n}=0.19(1)$ from the fitting shown in Fig.\,\ref{fig:rDetermination}. We confirm that $\sigma_{z,\mathrm{min}}$ converged at approximately $r=1.2$, which is consistent with theoretical value of $r=1.23$ that minimizes $\sigma_z|_{t_\mathrm{SP}=\pi/2\omega_1}$, obtained from Eq.\,\eqref{eq:sigmazatT1per4min}. However, when $r$ is of this magnitude, the position variance component becomes comparable to the momentum variance component ($\mathrm{e}^{4r}\simeq(\omega_0 t_\mathrm{TOF})^2\mathrm{e}^{-4r}$), making the interpretation of the reconstructed Wigner function no longer straightforward. In the main experiments, we set $r=0.890(4)$, where the impact of position variance is sufficiently negligible while providing adequate squeezing level.

\section{Estimation of background gas heating}

\begin{figure}
	\centering
	\includegraphics[width=\hsize]{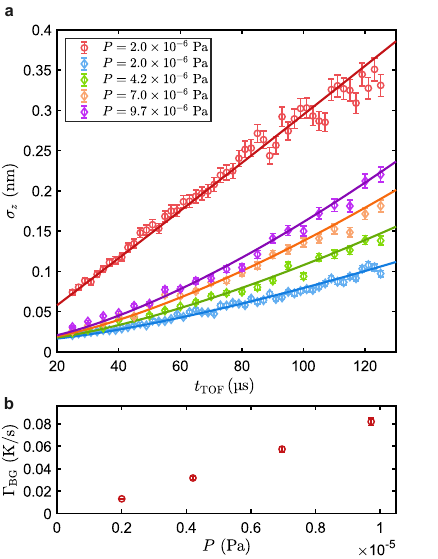}
	\caption{
    \textbf{Deterioration of $\sigma_z$ due to background gas heating.} 
    \textbf{a}, Dependence of $\sigma_z$ on $t_\mathrm{TOF}$ at various pressures. The circles (diamonds) show estimated $\sigma_z$ without (with) state preparation. Error bars show the standard error from Gaussian fits.
    \textbf{b}, Estimated $\Gamma_\mathrm{BG}$ for various pressures. Error bars show the standard error obtained in \textbf{a}.
    }
	\label{fig:BGHeating}
\end{figure}

We estimate the heating rate due to background gas by measuring the deterioration of momentum variance that becomes pronounced as $t_\mathrm{TOF}$ increases (Fig.\,\figref{fig:BGHeating}{a}). We first perform measurements without state preparation under the lowest possible pressure. We determine the occupation number $n$ from a fit without taking into account background gas heating. Using the obtained $n$, we estimated the heating rate due to background gas $\Gamma_\mathrm{BG}$ from the measurement obtained with state preparation performed at the same pressure. We then repeat the fit for the measurements without state preparation using the obtained $\Gamma_\mathrm{BG}$ and update $n$. This process is repeated until the estimated $\Gamma_\mathrm{BG}$ converges within $\unit[1]{\%}$. Using the resulting $n$ value, we estimate $\Gamma_\mathrm{BG}$ at various pressures from measurements obtained with state preparation (Fig.\,\figref{fig:BGHeating}{b}). From Eqs.\,\eqref{eq:VzzTOFwithSP} and \eqref{eq:VzzTOFBG}, the fitting function can be written as
\begin{equation}
    \sigma_z^\mathrm{BG}=\sqrt{\frac{\kappa\hbar}{2m\omega_0}\left[\left(\frac{\omega_0}{\omega_1}\right)^2+(\omega_1 t_\mathrm{TOF})^2\right]+\frac{2k_\mathrm{B}\Gamma_\mathrm{BG}}{3m}t_\mathrm{TOF}^3}.
\end{equation}
Here we performed the fitting with $\Gamma_\mathrm{BG}$ as the only free parameter. Because $t_\mathrm{SP}$ is not long compared to $t_\mathrm{TOF}$ and light power is approximately $28.5$ times smaller than that in cooling phase, heating during state preparation due to background gas and photon recoil can be neglected. The estimated results indicate that $\Gamma_\mathrm{BG}$ is proportional to pressure, consistent with theoretical predictions~\cite{PhysRevA.99.051401}. Furthermore, this result is consistent with our previous estimate of $\Gamma_\mathrm{BG}=\unit[16]{mK/s}$ at $\unit[3\times10^{-6}]{Pa}$~\cite{kamba2026levitatednanoaccelerometersensitizedquantum}. These results suggest that reducing pressure is crucial for extending the TOF duration to achieve better sensitivity. Currently, achievable pressures are limited because excessively low pressures cause particles to spin too rapidly and are lost.

\section{Validity of the TOF protocol as homodyne measurements}

We show that the measurement results obtained by the experimental sequence described in Fig.\,\figref{fig:setup}{b} correspond to the homodyne measurement results of the quadrature operator. The time evolution of the system during the experimental sequence can be described by two Hamiltonians: $\hat{H}_{\mathrm{(i)}}$ during the state preparation and $\hat{H}_{\mathrm{(ii)}}$ during the TOF expansion. The Hamiltonian $\hat{H}_{\mathrm{(i)}}$ is given by
\begin{align}
  \hat{H}_{\mathrm{(i)}}&=\frac{\hat{p}^2}{2m}+\frac{1}{2}m\omega_1^2\hat{z}^2-mg\sin\theta\hat{z} \\
  &=\frac{\hbar\omega_1}{4}\left[\hat{p}_1^2+\left(\hat{z}_1-\frac{g\sin\theta}{\omega_1}\sqrt{\frac{2m}{\hbar\omega_1}}\right)^2\right]+\mathrm{const.},
\end{align}
where $\hat{z}_1=\sqrt{\frac{2m\omega_1}{\hbar}}\hat{z}$ and $\hat{p}_1=\sqrt{\frac{2}{m\hbar\omega_1}}\hat{p}$ are the dimensionless position and momentum operators, with $\left[\hat{z}, \hat{p}\right]=i\hbar$. For simplicity, we introduce the shifted position operator $\hat{Z}_1=\hat{z}_1-\frac{g\sin\theta}{\omega_1}\sqrt{\frac{2m}{\hbar\omega_1}}$ to rewrite the Hamiltonian as
\begin{equation}
  \hat{H}_{\mathrm{(i)}}=\frac{\hbar\omega_1}{4}\left(\hat{p}_1^2+\hat{Z}_1^2\right)+\mathrm{const.}
\end{equation}
In the Heisenberg picture, the time evolution of the dimensionless shifted position and momentum operators $\hat{Z}_1, \ \hat{p}_1$ under the Hamiltonian $\hat{H}_{\mathrm{(i)}}$ is expressed as
\begin{align}
  \hat{Z}_1(t) &= \hat{Z}_1(0)\cos\left(\omega_1 t\right)+\hat{p}_1(0)\sin\left(\omega_1 t\right), \label{eq:Z1_evolution}\\
  \hat{p}_1(t) &= -\hat{Z}_1(0)\sin\left(\omega_1 t\right)+\hat{p}_1(0)\cos\left(\omega_1 t\right). \label{eq:p1_evolution}
\end{align}
The Hamiltonian $\hat{H}_{\mathrm{(ii)}}$ is given by
\begin{align}
  \hat{H}_{\mathrm{(ii)}}&=\frac{\hat{p}^2}{2m}-mg\sin\theta\hat{z} \\
  &=\frac{\hbar\omega_1}{4}\hat{p}_1^2-g\sin\theta\sqrt{\frac{\hbar m}{2\omega_1}}\hat{z}_1.
\end{align}
The time evolution of the dimensionless position operator $\hat{z}_1$ under the Hamiltonian $\hat{H}_{\mathrm{(ii)}}$ yields
\begin{equation}
  \hat{z}_1(t) = \hat{z}_1(t_\mathrm{SP})+\omega_1 t\hat{p}_1(t_\mathrm{SP})+\left(\omega_1 t\right)^2g\sin\theta\sqrt{\frac{m}{2\hbar\omega_1^3}}. \label{eq:z1_evolution}
\end{equation}
Using Eqs.\,\eqref{eq:Z1_evolution}, \eqref{eq:p1_evolution}, and \eqref{eq:z1_evolution}, the position operator $\hat{z}$ measured at time $t_{\mathrm{SP}}+t_{\mathrm{TOF}}$ under the initial condition $\hat{Z}_1(0)=\hat{Z}_1$ and $\hat{p}_1(0)=\hat{p}_1$ takes the form
\begin{align}
  &\hat{z}(t_{\mathrm{SP}}+t_{\mathrm{TOF}}) \notag\\
  &= \sqrt{\frac{\hbar}{2m\omega_1}}\left[\hat{Z}_1\cos\left(\omega_1 t_{\mathrm{SP}}\right)+\hat{p}_1\sin\left(\omega_1 t_{\mathrm{SP}}\right)\right. \notag\\
  &\quad\left.+\omega_1 t_{\mathrm{TOF}}\left(-\hat{Z}_1\sin\left(\omega_1 t_{\mathrm{SP}}\right)+\hat{p}_1\cos\left(\omega_1 t_{\mathrm{SP}}\right)\right)\right] \notag\\
  &\quad+\frac{g\sin\theta}{\omega_1^2}+\frac{1}{2}g\sin\theta t_{\mathrm{TOF}}^2 \\
  &=\sqrt{\frac{\hbar}{2m\omega_1}}\sqrt{1+\left(\omega_1 t_{\mathrm{TOF}}\right)^2}\hat{\tilde{p}}_1(\phi) \notag\\
  &\quad +\frac{g\sin\theta}{\omega_1^2}\left[1-\cos(\omega_1t_{\mathrm{SP}})+\omega_1 t_{\mathrm{TOF}}\sin(\omega_1t_{\mathrm{SP}})\right]\notag\\
  &\quad+\frac{1}{2}g\sin\theta t_{\mathrm{TOF}}^2,
\end{align}
where $\phi=\omega_1 t_{\mathrm{SP}}-\arctan\left(\frac{1}{\omega_1 t_{\mathrm{TOF}}}\right)$ is a rotated phase in the phase space and $\hat{\tilde{p}}_1(\phi)=-\hat{z}_1\sin\phi+\hat{p}_1\cos\phi$ is the rotated dimensionless momentum operator. Therefore, a projective measurement of the position at time $t_{\mathrm{SP}}+t_{\mathrm{TOF}}$ corresponds to the homodyne measurement of the quadrature $\hat{\tilde{p}}_1(\phi)$ with a certain offset. By varying the time $t_{\mathrm{SP}}$, we can measure the quadrature distribution at various phases $\phi$ in the phase space, which allows us to perform motional tomography of the mechanical state.

\section{Details of reconstructing the Wigner function}

Consider we have a set of homodyne measurement outcomes, $\{(\tilde{p}_i, \phi_i)\}_{i=1}^N$, where $\tilde{p}_i$ is the measured quadrature value at phase $\phi_i$ and $N$ is the total number of measurements. For any given pair $(\tilde{p}_i, \phi_i)$, we can define the projection operator onto the corresponding eigenstate of the quadrature, $\hat{\Pi}(\tilde{p}_i, \phi_i)=\ket{\tilde{p}_i, \phi_i}\!\bra{\tilde{p}_i, \phi_i}$. Then, with the system being in the quantum state $\hat{\rho}$, the likelihood can be expressed as
\begin{equation}
  \mathcal{L}_{\mathrm{point}}(\hat{\rho}) = \prod_{i=1}^N \mathrm{Tr}(\hat{\Pi}(\tilde{p}_i, \phi_i) \hat{\rho}). \label{eq:likelihood_point}
\end{equation}
To find the density matrix $\hat{\rho}$ which maximizes the likelihood in Eq.\,\eqref{eq:likelihood_point}, we need to construct a state $\hat{\rho}_0$ which satisfies the condition 
\begin{equation}
  \hat{R}_{\mathrm{point}}(\hat{\rho}_0)\hat{\rho}_0\hat{R}_{\mathrm{point}}(\hat{\rho}_0) = \hat{\rho}_0,
\end{equation}
where $\hat{R}_{\mathrm{point}}(\hat{\rho})$ is defined as
\begin{equation}
  \hat{R}_{\mathrm{point}}(\hat{\rho}) = \frac{1}{N}\sum_{i=1}^N \frac{\hat{\Pi}(\tilde{p}_i, \phi_i)}{\mathrm{Tr}(\hat{\Pi}(\tilde{p}_i, \phi_i) \hat{\rho})}. \label{eq:R_operator}
\end{equation}
Here, we introduce the diluted operator
\begin{equation}
  \hat{R}_{\mathrm{diluted}}(\hat{\rho}) = \hat{I}+\epsilon\left(\hat{R}_{\mathrm{point}}(\hat{\rho})-\hat{I}\right), \label{eq:R_diluted}
\end{equation}
where $\epsilon$ is a small positive constant, $0<\epsilon\leq1$. This diluted process improves the convergence of the iterative algorithm presented in the following. We may then form the state $\hat{\rho}_0$ by iteratively applying the operator $\hat{R}_{\mathrm{diluted}}(\hat{\rho})$ to an initial guess $\hat{\rho}^{(0)}=\mathcal{N}[\hat{I}]$ as follows:
\begin{equation}
  \hat{\rho}^{(k+1)} = \mathcal{N} \left[\hat{R}_{\mathrm{diluted}}(\hat{\rho}^{(k)})\hat{\rho}^{(k)}\hat{R}_{\mathrm{diluted}}(\hat{\rho}^{(k)})\right], \label{eq:iteration}
\end{equation}
where $k$ is the iteration step and $\mathcal{N}$ denotes the normalization to a unit trace. For sufficiently small $\epsilon$, the likelihood is guaranteed to be nondecreasing in every iteration step. This iterative process continues until convergence, which is typically determined by the change in the density matrix or the likelihood value between iterations falling below a specified threshold.\par

In practice, it is necessary to represent the density operator $\hat{\rho}$ as a matrix in a finite-dimensional Hilbert space for numerical calculations. For this purpose, we utilize the Fock basis $\{\ket{n}\}_{n=0}^{n_{\mathrm{max}}}$ and truncate the Hilbert space at a sufficiently large $n_{\mathrm{max}}$ to ensure that the reconstructed state is accurately represented. The projection operator $\hat{\Pi}(\tilde{p}_i, \phi_i)$ can be expressed in the Fock basis as
\begin{equation}
  \hat{\Pi}(\tilde{p}_i, \phi_i) = \sum_{m,n=0}^{n_{\mathrm{max}}} \braket{m|\tilde{p}_i, \phi_i}\!\braket{\tilde{p}_i, \phi_i|n} \ket{m}\!\bra{n}.
\end{equation}
The matrix elements $\Pi_{mn}(\tilde{p}_i, \phi_i) = \braket{m|\tilde{p}_i, \phi_i}\!\braket{\tilde{p}_i, \phi_i|n}$ can be calculated using the known solution of the Schr\"{o}dinger equation for a particle in a harmonic potential:
\begin{align}
  &\braket{m|\tilde{p}_i, \phi_i} \notag \\
  &= e^{i m(\phi_i+\pi/2)} \left(\frac{1}{2\pi}\right)^{1/4} \frac{H_m\left(\tilde{p}_i/\sqrt{2}\right)}{\sqrt{2^{m} m!}} e^{-\frac{1}{2}(\tilde{p}_i/\sqrt{2})^2},
\end{align}
where $H_m(x)$ is the $m$-th Hermite polynomial.\par

However, the iterative scheme mentioned above cannot be applied directly to the continuous data obtained from homodyne measurements. To address this issue, we first bin the homodyne measurement data into histograms with a finite bin width for each phase. We then apply the MLE algorithm to the binned data. The likelihood for the binned data can be expressed as
\begin{equation}
  \mathcal{L}_{\mathrm{bin}}(\hat{\rho}) = \prod_{i=1}^{N_{\phi}}\prod_{j=1}^{N_{\tilde{p}(\phi_i)}} \left[\mathrm{Tr}(\hat{\Pi}_{ij} \hat{\rho})\right]^{f_{ij}}, \label{eq:likelihood_bin}
\end{equation}
where $N_{\phi}$ is the total number of phases, $N_{\tilde{p}(\phi_i)}$ is the total number of bins at the phase $\phi_i$, and $f_{ij}$ is the frequency of the $(i,j)$-th bin. $\hat{\Pi}_{ij}$ is the projection operator corresponding to the $(i,j)$-th bin defined as
\begin{equation}
  \hat{\Pi}_{ij} = \int_{\tilde{p}_j-\Delta/2}^{\tilde{p}_j+\Delta/2} \mathrm{d}\tilde{p} \ket{\tilde{p}, \phi_i}\!\bra{\tilde{p}, \phi_i},
\end{equation}
where $\tilde{p}_j$ is the center of the $j$-th bin and $\Delta$ is the bin width. The operator $\hat{R}_{\mathrm{bin}}(\hat{\rho})$ for the binned data can be defined as
\begin{equation}
  \hat{R}_{\mathrm{bin}}(\hat{\rho}) = \frac{1}{N}\sum_{i=1}^{N_{\phi}}\sum_{j=1}^{N_{\tilde{p}(\phi_i)}} \frac{f_{ij} \hat{\Pi}_{ij}}{\mathrm{Tr}(\hat{\Pi}_{ij} \hat{\rho})}. \label{eq:R_bin}
\end{equation}
By replacing $\hat{R}_{\mathrm{point}}(\hat{\rho})$ with $\hat{R}_{\mathrm{bin}}(\hat{\rho})$ in the iteration formula Eq.\,\eqref{eq:iteration}, we can perform the MLE for the binned homodyne measurement data. The choice of the bin width $\Delta$ and the upper and lower bounds of the bins is crucial for the accuracy of the reconstructed state. Therefore, we have to set both bounds sufficiently wide to cover the range of the measurement data and choose the bin width $\Delta$ to be small enough to capture the details of the quadrature distribution while avoiding excessive statistical fluctuations in each bin. In our analysis, we choose the bin width to be 0.2 times the minimum of the standard deviations of the quadrature distributions at all phases.\par

We iteratively apply this procedure until convergence is achieved. The convergence is determined if a distance defined by the maximum norm,
\begin{equation}
  d(\rho^{(k)}, \rho^{(k+1)})=\max_{n,m}|\rho^{(k+1)}_{n,m}-\rho^{(k)}_{n,m}|,
\end{equation}
is less than a certain threshold and the relative change in the log likelihood,
\begin{equation}
  \left|\frac{\ln\mathcal{L}_{\mathrm{bin}}(\rho^{(k+1)})-\ln\mathcal{L}_{\mathrm{bin}}(\rho^{(k)})}{\ln\mathcal{L}_{\mathrm{bin}}(\rho^{(k)})}\right|,
\end{equation}
is less than another threshold between iterations in our analysis. In practical applications, we set the thresholds to be $3\times10^{-4}$ and $4\times10^{-5}$ for the thermal state, $9\times10^{-5}$ and $8\times10^{-6}$ for the thermal squeezed state respectively. We determine the thresholds by maximizing the fidelity between the reconstructed and true states using mock data. After convergence, validity of the truncation of the Hilbert space is checked by confirming that the population in the highest Fock state is negligible, $\rho_{n_{\mathrm{max}},n_{\mathrm{max}}}<10^{-4}$. To fulfill this condition, we set $n_{\mathrm{max}}=23$ for the thermal state, and we set $n_{\mathrm{max}}=70$ for the thermal squeezed state.\par

After obtaining the reconstructed density matrix $\hat{\rho}$ from the MLE algorithm, we calculate the Wigner function $W(z_1, p_1)$ of the reconstructed state using the formula
\begin{align}
  &W(z_1, p_1) \notag\\
  &= \frac{1}{2\pi} e^{-2|\alpha|^2} \sum_{n,m=0}^{n_{\mathrm{max}}} \rho_{nm} (-1)^{l} e^{-i\lambda(m-n)} \notag \\
  &\hspace{10em}\sqrt{\frac{l!}{(l+\delta)!}} |2\alpha|^{\delta} L_l^{\delta}(4|\alpha|^2),
\end{align}
where $\alpha = (z_1 + i p_1)/2$, $\lambda=\arg(\alpha)$, $l=\min(n,m)$, $\delta=|m-n|$ and $L_l^{\delta}(x)$ is the generalized Laguerre polynomial.\par 

\section{Estimation of the standard deviation of the reconstructed state}

To estimate the standard deviation of the reconstructed Wigner function, we perform another MLE procedure assuming a Gaussian model for the Wigner function. The Gaussian model is defined as
\begin{equation}
  W_{\mathrm{G}}(\bm{v}) = \frac{1}{2\pi\sqrt{\det(\Sigma)}} \exp\left[-\frac{1}{2}(\bm{v}-\bm{\mu})^{\top} \Sigma^{-1} (\bm{v}-\bm{\mu})\right],
\end{equation}
where $\bm{v}=(z_1,p_1)^\top$ is a vector of the position and momentum quadratures, $\bm{\mu}=(\mu_{z_1},\mu_{p_1})^\top$ is a mean vector, and $\Sigma$ is the covariance matrix defined as
\begin{equation}
  \Sigma = \begin{pmatrix} A-B_{\mathrm{c}} & -B_{\mathrm{s}} \\ -B_{\mathrm{s}} & A+B_{\mathrm{c}} \end{pmatrix},
\end{equation}
where $A$, $B_{\mathrm{c}}$ and $B_{\mathrm{s}}$ are parameters to be estimated. The parameters of the Gaussian model $\bm{\Theta}=(\mu_{z_1}, \mu_{p_1}, A, B_{\mathrm{c}}, B_{\mathrm{s}})$, can be estimated by maximizing the likelihood function defined as
\begin{equation}
  \mathcal{L}_{\mathrm{G}}(\bm{\Theta}) = \prod_{i=1}^{N} \frac{1}{\sqrt{2\pi V(\bm{\Theta}, \phi_i)}} \exp\left[-\frac{\left(\tilde{p}_i-M(\bm{\Theta}, \phi_i)\right)^2}{2V(\bm{\Theta}, \phi_i)}\right], \label{eq:likelihood_gaussian}
\end{equation}
where $M(\bm{\Theta}, \phi_i)$ and $V(\bm{\Theta}, \phi_i)$ are the estimated mean and variance for the $i$-th data point defined as follows:
\begin{align}
  M(\bm{\Theta}, \phi_i) &= -\mu_{z_1}\sin\phi_i+\mu_{p_1}\cos\phi_i, \\
  V(\bm{\Theta}, \phi_i) &= A+B_{\mathrm{c}}\cos(2\phi_i)+B_{\mathrm{s}}\sin(2\phi_i).
\end{align}
The maximization of the likelihood function can be performed using an interior-point method under the constraint $A>\sqrt{B_{\mathrm{c}}^2+B_{\mathrm{s}}^2}$, which ensures the covariance matrix is positive definite. The relationship between the estimated parameters and the dimensionless standard deviations of the state can be expressed as
\begin{equation}
  \sigma_{\pm} = \sqrt{A\pm\sqrt{B_{\mathrm{c}}^2+B_{\mathrm{s}}^2}},
\end{equation}
where the subscript $\pm$ corresponds to the antisqueezed and squeezed axes of the Gaussian distribution, respectively.\par

\begin{figure}
	\centering
	\includegraphics[width=\hsize]{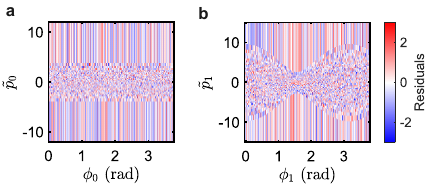}
	\caption{
    \textbf{Lumped sinograms and Poisson deviance residuals for the G-test.} 
    \textbf{a},\textbf{b}, Lumped sinograms of observed homodyne quadratures and Poisson deviance residuals for the thermal state and the thermal squeezed state, respectively. The bins in the original sinogram where the expected count is less than $10$ are merged deterministically. $\tilde{p}_0$ and $\tilde{p}_1$ are the quadrature values normalized by the zero-point fluctuation at the corresponding frequency, $\omega_0$ and $\omega_1$, respectively.
    }
	\label{fig:statplot}
\end{figure}

Following the estimation of model parameters, we perform a G-test on the sinogram residuals to assess the validity of the Gaussian ansatz. If the data follow a Gaussian model, the counts in each bin should follow a Poisson distribution characterized by the expected count of the model. The G-test is a likelihood ratio test that takes this assumption as a null hypothesis. It is known that the statistic used in the G-test (hereafter referred to as G-statistic) asymptotically follows a $\chi^2$ distribution if the null hypothesis holds. To satisfy this asymptotic property, we perform adaptive lumping, a procedure that deterministically merges bins in the original sinogram where the expected count is less than $10$. Figure\,\ref{fig:statplot} shows lumped sinograms and the corresponding Poisson deviance residuals interpreted as the extent to which each bin contributes to the overall deviance (G-statistic). Since the residuals are randomly distributed regardless of angle or homodyne measurement, it is evident that there is no systematic bias in the model.\par

\begin{table*}
    \centering
    \normalsize
    \caption{\textbf{Results of the G-test for the Gaussian model.} The G-statistic, degrees of freedom, upper p-value and corresponding percentile are shown for both the thermal state and the thermal squeezed state. The upper p-value indicates the probability of observing a G-statistic as extreme as, or more extreme than, the one calculated from the data under the null hypothesis. The percentile is calculated from the upper p-value, by converting it to a z-score using the inverse cumulative distribution function of the standard normal distribution.}
        \begin{tabular}{ccccc}
            \hline
            Type of state & G-statistic & Degrees of freedom & Upper p-value & Percentile\\
            \hline \hline
            Thermal state & 9286.86 & 9295 & $5.22\times10^{-1}$ & $-0.05\sigma$ \\
            Thermal squeezed state & 11925.26 & 11792 & $1.92\times10^{-1}$ & $+0.87\sigma$ \\
            \hline
        \end{tabular}
    \label{tab:gtest}
\end{table*}

We then conduct the G-test with the Williams' correction~\cite{Williams1976}. The results of the G-test are shown in Table\,\ref{tab:gtest}. For both states, the percentile of the data is within 1-sigma interval. This means our model provides a good fit to the experimental data. Our results demonstrate a high degree of consistency between the model and the experimental data, which justifies the use of the estimated standard deviation as a definitive measure of the state's phase-space distribution.\par

\begin{figure}
	\centering
	\includegraphics[width=\hsize]{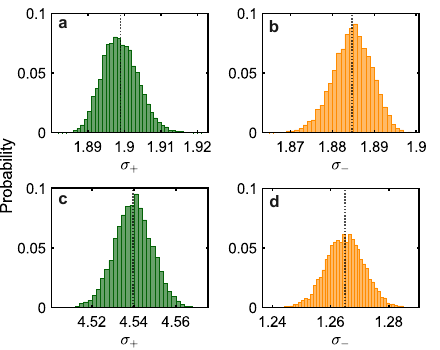}
	\caption{
    \textbf{Posterior distributions of the dimensionless standard deviations.} 
    \textbf{a}--\textbf{d}, Posterior distributions of $\sigma_\pm$ for the thermal state (\textbf{a},\textbf{b}) and the thermal squeezed state (\textbf{c},\textbf{d}). The horizontal axis represents the standard deviation normalized by the zero-point fluctuation at the corresponding frequency: $\omega_0$ for the thermal state and $\omega_1$ for the thermal squeezed state. The dashed vertical line indicates the mean value of each distribution.
    }
	\label{fig:mcmchistoplot}
\end{figure}

After the model is verified to be valid, we evaluate the uncertainty of the estimated standard deviation using Markov chain Monte Carlo (MCMC) sampling. We use the Metropolis-Hastings algorithm to generate samples from the posterior distribution of the parameters given the data, which is proportional to the likelihood function in Eq.\,\eqref{eq:likelihood_gaussian}. The MCMC sampling allows us to obtain a distribution of parameter values that are consistent with the observed data, and we can use this distribution to calculate credible intervals for the estimated standard deviation of the state. To ensure the robustness and independence of the posterior samples, we apply rigorous diagnostic criteria for burn-in and thinning. Specifically, we determine the burn-in period dynamically by evaluating the rank-normalized split-$\hat{R}$ statistic across multiple independent chains~\cite{27873dd5a58a4f5592ed70447736df1b}. Then, we discard the initial portion of the chains until $\hat{R}<1.01$ is satisfied for all parameters, after which we treat the remaining samples as having converged. Subsequently, to mitigate autocorrelation and extract robust samples, we thin the post-burn-in chains based on the integrated autocorrelation time (IACT). The IACT is estimated using Geyer's initial monotone sequence (IMS) estimator~\cite{10.1214/ss/1177011137}. This comprehensive diagnostic procedure yields a robust set of samples for reliable uncertainty quantification. In our implementation, $8$ chains are run in parallel and around $6000$ effective samples are obtained. Figure\,\ref{fig:mcmchistoplot} shows the posterior distributions of the dimensionless standard deviations $\sigma_{\pm}$ for both the thermal state and the thermal squeezed state. For the thermal state, we take the mean of $\sigma_{+}$ and $\sigma_-$ as the estimated standard deviation. We define the total uncertainty as the central $\unit[68]{\%}$ interval of each posterior distribution.

\section{Derivation of the Fisher information}

Here, we calculate the classical Fisher information from the density matrices reconstructed by quantum state tomography. We assume $\omega_0 \gg \omega_1$ and neglect the $\theta$-dependence of the initial thermal-equilibrium position in the $\omega_0$ trap, whose displacement is of order $g\sin\theta/\omega_0^2$. Under this approximation, the initial state can be treated as $\theta$-independent.\par

We first consider the Fisher information for the state-prepared protocol. In this case, the trap frequency is reduced from $\omega_0$ to $\omega_1$ and kept at $\omega_1$ for a duration $t_{\mathrm{SP}}$ before the TOF expansion. During the state-preparation interval, $0 \le t \le t_{\mathrm{SP}}$, the position and momentum operators evolve as
\begin{align}
\hat z(t) =& \cos(\omega_1 t)\,\hat z(0) + \frac{\sin(\omega_1 t)}{m\omega_1}\,\hat p(0) + g\theta\,\frac{1-\cos(\omega_1 t)}{\omega_1^2}, \\
\hat p(t) =& -\,m\omega_1\sin(\omega_1 t)\,\hat z(0) + \cos(\omega_1 t)\,\hat p(0) \nonumber\\
& + mg\theta\,\frac{\sin(\omega_1 t)}{\omega_1}.
\end{align}
During the TOF interval,
\begin{align}
&\hat z(t_{\mathrm{SP}}+t_{\mathrm{TOF}}) \nonumber\\
=& \hat z(t_{\mathrm{SP}}) + \frac{t_{\mathrm{TOF}}}{m}\hat p(t_{\mathrm{SP}}) + \frac{1}{2} g\theta\,t_{\mathrm{TOF}}^2\nonumber\\
=& \left( \cos\omega_1 t_{\mathrm{SP}} - \omega_1 t_{\mathrm{TOF}}\sin\omega_1 t_{\mathrm{SP}} \right)\hat z(0) \nonumber\\
& + \left( \frac{\sin\omega_1 t_{\mathrm{SP}}}{m\omega_1} + \frac{t_{\mathrm{TOF}}\cos\omega_1 t_{\mathrm{SP}}}{m} \right)\hat p(0) \nonumber\\
& + g\theta \left( \frac{1-\cos\omega_1 t_{\mathrm{SP}}}{\omega_1^2} + \frac{t_{\mathrm{TOF}}\sin\omega_1 t_{\mathrm{SP}}}{\omega_1} + \frac{t_{\mathrm{TOF}}^2}{2} \right).
\end{align}
Here, we use the small-angle approximation and set $\sin\theta \simeq \theta$.\par

In the experimentally relevant regime of sufficiently long TOF ($\omega_1 t_\mathrm{TOF}\gg 1$), we neglect the release-position contribution in the final-position operator and obtain
\begin{align}
\hat z(t_{\mathrm{SP}}+t_{\mathrm{TOF}}) \simeq \frac{t_{\mathrm{TOF}}}{m}\left.\hat p(t_{\mathrm{SP}})\right|_{\theta=0} + D_{\theta}\,\theta,
\end{align}
with
\begin{align}
D_{\theta} = \frac{g t_{\mathrm{TOF}}\sin\omega_1 t_{\mathrm{SP}}}{\omega_1} + \frac{g t_{\mathrm{TOF}}^2}{2}.
\end{align}
The Fisher information associated with a position measurement~\cite{PhysRevLett.72.3439} is
\begin{align}
F_z = \int \mathrm{d}z\, \frac{\left[\partial_\theta P_z(z|\theta)\right]^2}{P_z(z|\theta)},
\end{align}
with
\begin{align}
P_z(z|\theta) = \mathrm{Tr}\!\left[ \hat{\rho}\, \delta\!\left(z-\hat z(t_{\mathrm{SP}}+t_{\mathrm{TOF}})\right) \right].
\end{align}
Accordingly, the final-position distribution is approximated as a rigidly shifted and rescaled version of the distribution of $\left.\hat p(t_{\mathrm{SP}})\right|_{\theta=0}$:
\begin{align}
P_z(z|\theta) \simeq \frac{m}{t_{\mathrm{TOF}}}\, P_{\left.p(t_{\mathrm{SP}})\right|_{\theta=0}}\!\left( \frac{m}{t_{\mathrm{TOF}}} \left[z-D_{\theta}\theta\right] \right),
\end{align}
where $P_{\left.p(t_{\mathrm{SP}})\right|_{\theta=0}}$ is the probability density of $\left.\hat p(t_{\mathrm{SP}})\right|_{\theta=0}$.\par

The Fisher information for a position measurement in the long-TOF approximation is then
\begin{align}
F_z \simeq \frac{D_{\theta}^2}{(t_{\mathrm{TOF}}/m)^2} \,I\!\left[P_{\left.p(t_{\mathrm{SP}})\right|_{\theta=0}}\right],
\end{align}
where\par
\begin{align}
I[P] = \int \mathrm{d}q\, \frac{\left[\partial_q P(q)\right]^2}{P(q)}
\end{align}
denotes the classical Fisher information associated with translations of the probability density $P(q)$.\par

Equivalently, we define the dimensionless operators normalized by the zero-point fluctuations at frequency $\omega_1$ as
\begin{align}
\hat z_1 &= \frac{\hat z(0)}{\sqrt{\hbar/(2m\omega_1)}}, \qquad \hat p_1 = \frac{\hat p(0)}{\sqrt{\hbar m\omega_1/2}},
\end{align}
and introduce the rotated quadrature
\begin{align}
\hat{\tilde p}_1(\phi_1) = -\hat z_1\sin\phi_1+\hat p_1\cos\phi_1.
\end{align}
Then
\begin{align}
\left.\hat p(t_{\mathrm{SP}})\right|_{\theta=0} = \sqrt{\frac{\hbar m\omega_1}{2}}\, \hat{\tilde p}_1(\phi_1), \qquad \phi_1=\omega_1 t_{\mathrm{SP}}.
\end{align}
Therefore,
\begin{align}
F_z \simeq \frac{D_{\theta}^2}{(t_{\mathrm{TOF}}/m)^2\,(\hbar m\omega_1/2)} I[P_{\phi_1}],
\end{align}
where $P_{\phi_1}$ is the probability density of $\hat{\tilde p}_1(\phi_1)$.\par

We next consider the case without state preparation. In the absence of state preparation, the trap frequency is not reduced and remains $\omega_0$ during the interval $0 \le t \le t_{\mathrm{SP}}$. As in the case with state preparation, we begin from the operator evolution and then evaluate the Fisher information associated with a position measurement after TOF. To distinguish the two cases, we denote the resulting quantities without state preparation by the superscript $(\mathrm{w/o})$. During this interval, the position and momentum operators evolve as
\begin{align}
\hat z(t) =& \cos(\omega_0 t)\,\hat z(0) + \frac{\sin(\omega_0 t)}{m\omega_0}\,\hat p(0) + g\theta\,\frac{1-\cos(\omega_0 t)}{\omega_0^2}, \\
\hat p(t) =& -\,m\omega_0\sin(\omega_0 t)\,\hat z(0) + \cos(\omega_0 t)\,\hat p(0) \nonumber\\
&+ mg\theta\,\frac{\sin(\omega_0 t)}{\omega_0}.
\end{align}
During the TOF interval,
\begin{align}
&\hat z(t_{\mathrm{SP}}+t_{\mathrm{TOF}}) \nonumber\\
= & \hat z(t_{\mathrm{SP}}) + \frac{t_{\mathrm{TOF}}}{m}\hat p(t_{\mathrm{SP}}) + \frac{1}{2} g\theta\,t_{\mathrm{TOF}}^2 \nonumber\\
= & \left( \cos\omega_0 t_{\mathrm{SP}} - \omega_0 t_{\mathrm{TOF}}\sin\omega_0 t_{\mathrm{SP}} \right)\hat z(0) \nonumber\\
&+ \left( \frac{\sin\omega_0 t_{\mathrm{SP}}}{m\omega_0} + \frac{t_{\mathrm{TOF}}\cos\omega_0 t_{\mathrm{SP}}}{m} \right)\hat p(0) \nonumber\\
&+ g\theta \left( \frac{1-\cos\omega_0 t_{\mathrm{SP}}}{\omega_0^2} + \frac{t_{\mathrm{TOF}}\sin\omega_0 t_{\mathrm{SP}}}{\omega_0} + \frac{t_{\mathrm{TOF}}^2}{2} \right).
\end{align}
The Fisher information is defined analogously to the state-prepared case. In the experimentally relevant regime of sufficiently long TOF, satisfying $\omega_0 t_{\mathrm{TOF}} \gg 1$, the release-position contribution in the final-position operator is negligible compared with the TOF-mapped momentum contribution.\par

We neglect the release-position contribution in the final-position operator and obtain
\begin{align}
\hat z(t_{\mathrm{SP}}+t_{\mathrm{TOF}}) \simeq \frac{t_{\mathrm{TOF}}}{m}\left.\hat p(t_{\mathrm{SP}})\right|_{\theta=0} + D_\theta^{(\mathrm{w/o})}\,\theta,
\end{align}
with
\begin{align}
D_\theta^{(\mathrm{w/o})} = \frac{g t_{\mathrm{TOF}}\sin\omega_0 t_{\mathrm{SP}}}{\omega_0} + \frac{g t_{\mathrm{TOF}}^2}{2}.
\end{align}
The Fisher information for a position measurement in the long-TOF approximation is
\begin{align}
F_z^{(\mathrm{w/o})} \simeq
\frac{\left(D_\theta^{(\mathrm{w/o})}\right)^2}
{(t_{\mathrm{TOF}}/m)^2}
\,I\!\left[P_{\left.p(t_{\mathrm{SP}})\right|_{\theta=0}}^{(\mathrm{w/o})}\right],
\end{align}
where $P_{\left.p(t_{\mathrm{SP}})\right|_{\theta=0}}^{(\mathrm{w/o})}$ is the probability density of $\left.\hat p(t_{\mathrm{SP}})\right|_{\theta=0}$.\par

We define the dimensionless operators normalized by the zero-point fluctuations at frequency $\omega_0$ as
\begin{align}
\hat z_0 &= \frac{\hat z(0)}{\sqrt{\hbar/(2m\omega_0)}}, \qquad
\hat p_0 = \frac{\hat p(0)}{\sqrt{\hbar m\omega_0/2}},
\end{align}
and introduce the rotated quadrature
\begin{align}
\hat{\tilde p}_0(\phi_0) = -\hat z_0\sin\phi_0+\hat p_0\cos\phi_0.
\end{align}
Then
\begin{align}
\left.\hat p(t_{\mathrm{SP}})\right|_{\theta=0}
= \sqrt{\frac{\hbar m\omega_0}{2}}\,\hat{\tilde p}_0(\phi_0),
\qquad \phi_0=\omega_0 t_{\mathrm{SP}}.
\end{align}
Therefore, the Fisher information can be written as
\begin{align}
F_z^{(\mathrm{w/o})} \simeq
\frac{\left(D_\theta^{(\mathrm{w/o})}\right)^2}
{(t_{\mathrm{TOF}}/m)^2\,(\hbar m\omega_0/2)}
\,I\!\left[P_{\phi_0}^{(\mathrm{w/o})}\right],
\end{align}
where $P_{\phi_0}^{(\mathrm{w/o})}$ is the probability density of $\hat{\tilde p}_0(\phi_0)$.\par

Finally, to estimate the statistical uncertainty of the Fisher information, we generate an ensemble of tomographically reconstructed density matrices by bootstrap resampling of the experimental data and evaluate the Fisher information for each sample using the same numerical procedure as described above. The quoted value is taken as the bootstrap mean, and the uncertainty is defined by the central $\unit[68]{\%}$ interval of the resulting bootstrap distribution.\par

\end{document}